# A Scaling Approach to Evaluating the Distance Exponent of the Urban Gravity Model

Yanguang Chen, Linshan Huang

(Department of Geography, College of Urban and Environmental Sciences, Peking University, Beijing 100871, P.R. China. E-mail: chenyg@pku.edu.cn)

**Abstract**: The gravity model is one of important models of social physics and human geography, but several basic theoretical and methodological problems remain to be solved. In particular, it is hard to explain and evaluate the distance exponent using the ideas from Euclidean geometry. This paper is devoted to exploring the distance-decay parameter of the urban gravity model. Based on the concepts from fractal geometry, several fractal parameter relations can be derived from scaling laws of self-similar hierarchies of cities. Results show that the distance exponent is just a scaling exponent, which equals the average fractal dimension of the size measurements of the cities within a geographical region. The scaling exponent can be evaluated with the product of Zipf's exponent of size distributions and the fractal dimension of spatial distributions of geographical elements such as cities and towns. The new equations are applied to China's cities, and the empirical results accord with the theoretical expectations. The findings lend further support to the suggestion that the geographical gravity model is a fractal model, and its distance exponent is associated with a fractal dimension and Zipf's exponent. This work will help geographers understand the gravity model using fractal theory and estimate the distance exponent using fractal modeling.

# 1 Introduction

A set of gravity models have been applied to explain and predict various behaviors of spatial interactions in many social sciences. Among this family of models, the basic one is the gravity model of migration based on an inverse power-law distance-decay effect, which was proposed by analogy with Newton's law of gravitation. It can be used to describe the strength of interaction between two places (Haynes and Fotheringham, 1984; Liu and Sui *et al*, 2014; Rodrigue *et al*, 2009; Sen and Smith, 1995). According to the model, any two places attract each another by a force that is directly proportional to the product of their sizes and inversely proportional to the *b*th power of the distance between them, and *b* is the distance-decay exponent, which is often termed *distance exponent* for short (Haggett *et al*, 1977). The size of a place can be measured with appropriate variables such as population numbers, built-up area, and gross domestic product (GDP). Recently, the gravity model has been employed to study the attractive effect of various new-fashioned human and physical activities by means of modern technology (Balcan *et al*, 2009; Goh *et al*, 2012; Kang *et al*, 2012; Kang *et al*, 2013; Jung *et al*, 2008; Lee *et al*, 2014; Liang, 2009; Liu and Wang *et al*, 2014; Simini *et al*, 2012). The model is empirically effective for describing spatial interactions, but it is obstructed by two problems on the distance exponent, *b*. One is the dimensional problem, that is, it is impossible to interpret the distance exponent in light of Euclidean geometry (Haggett *et al*, 1977; Haynes, 1975). The other is the algorithmic problem, namely, it is hard to estimate the numerical value of the distance exponent (Mikkonen and Luoma, 1999).

The first problem can be readily solved by using ideas from fractals. Fractal geometry was developed by Mandelbrot (1982), and it is a powerful tool for spatial analysis in geographical research (Batty, 2005; Batty and Longley, 1994; Frankhauser, 1994; Frankhauser, 1998). It can be proved that the distance exponent is a fractal parameter indicating a space dimension (Chen, 2015). However, how to evaluate the distance exponent is still a difficult problem that remains to be solved. A new discovery is that the distance exponent is associated with the Zipf exponent of the rank-size distribution and the fractal dimension of a self-organized network of urban places (Chen, 2011). If we examine the spatial interaction between cities and towns in a region, the distance exponent of the gravity model is just the product of the Zipf exponent of the city-size distribution and the fractal dimension of the corresponding central-place network. The formula has been derived by combining central-place theory and the rank-size rule (Chen, 2011). A pending problem is how to use the formula to estimate the distance exponent for gravity analysis in empirical studies.

This problem can be solved using a self-similar hierarchy with cascade structure. On the one hand, the rank-size distribution is mathematically equivalent to the hierarchical structure (Chen, 2012a; Chen, 2012b). On the other, the hierarchical structure and the network structure represent two different sides of the same coin (Batty and Longley, 1994). In short, the rank-size distribution and the network structure can be linked with one another by hierarchical structure. Thus, the Zipf exponent of the rank-size distribution and the fractal dimension of spatial network can be associated with each other through hierarchical scaling, which suggests a new approach to estimating the distance exponent. This study is mainly based on urban geography and is devoted to exploring the methods for distance exponent estimation. The rest of this article is organized as follows. In Section 2, a scaling relation will be derived from the gravity model based on the inverse power law, and spatial scaling will be transformed into hierarchical scaling in terms of the inherent association between hierarchy and network. Three scaling approaches to estimating the distance exponent will be proposed. In Section 3, Chinese cities will be employed to make an empirical analysis by means of census data and spatial distance data. In Section 4, several related questions will be discussed. Finally, the discussion will be concluded by summarizing the main points in the work.

## 2 Theoretical results

### 2.1 Breaking-point relation

The new method can be derived from the well-known breaking-point formula based on the basic gravity model. Consider three locations within a geographical region, *i*, *j*, and *x*, and *x* falls between *i* and *j*. If there are cities or towns at these locations, the "mass" of the settlements can be measured by population or other size quantity. Thus the gravity can be expressed as

$$I_{ix} = G \frac{P_i P_x}{L_{ix}^b}, \tag{1}$$

$$I_{jx} = G \frac{P_j P_x}{L_{jx}^b}, \tag{2}$$

where $I_{ix}$ denotes the gravity between locations *i* and *x*, $I_{jx}$ denotes the gravity between location *j*

and $x$, $L_{ix}$ indicates the distance between locations $i$ and $x$, $L_{jx}$ indicates the distance between locations $j$ and $x$, $P_i$, $P_j$, and $P_x$ are the size of the cities at locations $i$, $j$, and $x$, $G$ refers to the gravity coefficient, and $b$ to the distance exponent indicative of spatial friction. Combining equations (1) and (2) yields

$$\frac{P_i / L_i^b}{P_j / L_j^b} = \frac{I_{ix}}{I_{jx}}. \tag{3}$$

Suppose that there exists a special location for $x$, where $I_{ix}=I_{jx}$. In this case, eliminating $x$ yields

$$\frac{P_i}{P_j} = (\frac{L_{ix}}{L_{jx}})^b, \tag{4}$$

which is familiar to geographers. Chen (2015) demonstrated that equation (4) represents a fractal dimension relationship. Equation (4) is identical in form to the well-known breaking-point formula derived by Reilly (1931) and Converse (1949). By using the hierarchical scaling laws of urban systems, we can reveal the spatial meaning of the parameter $b$ and find a new way of evaluating it.

**2.2 Distance exponent based on hierarchical scaling laws**

If a geographical region is large enough, the size distribution of cities within the region may be consistent with Zipf's law. However, the laws of complex social and economic systems are not of spatio-temporal translational symmetry, and city-size distributions do not follow the only law (Chen, 2016). Sometimes, Zipf's law is replaced by other mathematical laws such as Lavalette law (Cerqueti and Ausloos, 2015). Zipf's law is one of the well-known rank-size scaling law, indicating self-organized criticality of urban evolution (Bak, 1996; Chen and Zhou, 2008). Suppose that city development in a region fall into self-organized critical state. The general Zipf formula can be expressed as follows (Zipf, 1949)

$$P(k) = P_1 k^{-q}, \tag{5}$$

where $k$ refers to the rank of cities in descending order, $P(k)$ to the size of the city of rank $k$, $q$ denotes the Zipf scaling exponent, and the proportionality coefficient $P_1$ indicates the size of the largest city in theory. City size is always measured with resident population. The inverse function of equation (5) is $k = [P(k)/P_1]^{-1/q}$, in which the rank $k$ represents the number of cities with size greater than or equal to $P(k)$. Reducing $P(k)$ to $P$ and substituting $k$ with $N(P)$, we have

$$N(P) = \eta P^{-p}, \tag{6}$$

which can be converted into the function of Pareto's density distribution. This implies that Zipf's law is equivalent in mathematics to Pareto's law and that the Zipf distribution is theoretically equivalent to the Pareto distribution (Chen, 2014a). In equation (6), the coefficient $\eta = P_1^p$ refers to the proportionality constant, and the power exponent $p=1/q$ denotes the Pareto scaling exponent, indicating the fractal dimension of city rank-size distributions. The reciprocal relation between $p$ and $q$ is based on pure mathematical derivation. In empirical analyses, this relation proved to be replaced by $p=R^2/q$, where $R$ denotes Pearson correlation coefficient (Chen, 2016; Chen and Zhou, 2003; Tan and Fan, 2004).

The rank-size distribution is actually a signature of the self-organized hierarchy with cascade structure. It has been demonstrated that if the city-size distribution follows Zipf's law, the cities can be organized into a self-similar hierarchy (Chen, 2012a; Chen, 2012b). The hierarchy of cities consisting of $M$ classes (levels) in a top-down order can be described with the discrete expressions of two exponential functions as below:

$$N_m = N_1 r_n^{m-1}, \tag{7}$$

$$P_m = P_1 r_p^{1-m}, \tag{8}$$

where $m=1, 2, \ldots, M$ refers to the order of classes in the hierarchy of cities ($M$ is a positive integer), $N_m$ and $P_m$ denote the number of cities and average size of the cities in the $m$th class, $N_1$ and $P_1$ are the number and mean size of the cities in the top class, and $r_n = N_{m+1}/N_m$ and $r_p = P_m/P_{m+1}$ are the *number ratio* and *size ratio* of cities, respectively. If $r_f=2$ as given, then the value of $r_p$ can be calculated; if $r_p=2$ as given, the value of $r_p$ can be derived (Chen, 2012b; Davis, 1978). According to the central-place theory propounded by Christaller (1933/1966), we have the third exponential model such as (Chen, 2011)

$$L_m = L_1 r_l^{1-m}, \tag{9}$$

where $L_m$ denotes the average distance between two urban places in the $m$th class, $L_1$ is the average distance between the urban places in the top class, and $r_l = L_m/L_{m+1}$ is the *distance ratio* of cities (the subscript of $L$ is the number 1, and the subscript of $r$ is the letter $l$). Here we assume there more than two cities at the first level of an urban hierarchy. If there are just two cities in the first class, $L_1$ will refer to the distance between the two cities; if there is only one top-level city, $L_1$ can be substituted by the radius of the equivalent circle of a study area.

The fractal model and hierarchical scaling relations can be derived from the three exponential functions. From equations (7) and (8), the size-number hierarchical scaling relation can be derived as

$$N_m = \mu P_m^{-p}, \tag{10}$$

which is equivalent to Pareto's law, equation (6) [see Appendix 1]. The proportionality coefficient is $\mu = N_1 P_1^p$, and the scaling exponent can be redefined as

$$p = -\frac{\ln(N_m / N_{m+1})}{\ln(P_m / P_{m+1})} = \frac{\ln r_n}{\ln r_p}, \tag{11}$$

which is regarded as the similarity dimension of city-size distributions. It is in fact a ratio of two fractal dimensions (Chen, 2014a). From equations (7) and (9), a fractal model of central places is derived as

$$N_m = \pi L_m^{-D}, \tag{12}$$

where $\pi = N_1 L_1^D$ refers to the proportionality coefficient, and $D$ to the fractal dimension of spatial distribution of urban places. The similarity dimension of fractals can be given by

$$D = -\frac{\ln(N_m / N_{m+1})}{\ln(L_m / L_{m+1})} = \frac{\ln r_n}{\ln r_l}, \tag{13}$$

which is the fractal parameter of central place networks and can be termed the *network dimension*. From equations (8) and (9), a size-range allometric relation is derived as

$$P_m = \kappa L_m^\sigma, \tag{14}$$

where $\kappa = P_1 L_1^{-\sigma}$ refers to the proportionality coefficient and $\sigma$ to the allometric scaling exponent of population distribution. The allometric exponent can be given by

$$\sigma = \frac{\ln(P_m / P_{m+1})}{\ln(L_m / L_{m+1})} = \frac{\ln r_p}{\ln r_l}, \tag{15}$$

which suggests that the allometric exponent is actually a fractal dimension of urban population.

Using the theoretical framework of urban hierarchical scaling, we can derive a new equation for the distance exponent of the gravity model. Based on the hierarchical structure of urban systems, the breaking-point relation can be rewritten as

$$\frac{P_m}{P_{m+1}} = \left(\frac{L_m}{L_{m+1}}\right)^b, \tag{16}$$

which is derived from equation (4) and can be regarded as a variant of equation (4) [see Appendix 2]. Since $r_p = P_m/P_{m+1}$, and $r_l = L_m/L_{m+1}$, equation (16) can be expressed in the form

$$r_p = r_l^b. \tag{17}$$

Thus we have

$$b = \frac{\ln(P_m/P_{m+1})}{\ln(L_m/L_{m+1})} = \frac{\ln r_p}{\ln r_l} = \sigma. \tag{18}$$

This suggests that the distance exponent is equal to the allometric scaling exponent of city size and its spatial distribution. By equations (11), (13), and (15), we can get

$$D = \frac{\ln r_n}{\ln r_l} = \frac{\ln r_p}{\ln r_l}\frac{\ln r_n}{\ln r_p} = \sigma p = \frac{b}{q}, \tag{19}$$

which suggests

$$b = qD. \tag{20}$$

For the classical central-place systems, $D \to 2$. Thus, if $q \to 1$, we will have $b \to 2$. The special value $b=2$ is used in empirical analyses (Jung *et al*, 2008). The Zipf exponent can be calculated by rank-size scaling analysis, and the network dimension can be computed by self-similar network analysis. Therefore, the distance exponent can be readily estimated by the parameter relationship shown in equation (20).

## 2.3 Three approaches for evaluating the distance exponent

New methods can be found for estimating the distance exponent of the urban gravity model. The Pareto scaling exponent has been proved to be a ratio of the fractal dimension of a network of cities to the average dimension of city population (Chen, 2014a). Accordingly, the Zipf scaling exponent is the reciprocal of this dimension ratio, that is

$$q = \frac{1}{p} = \frac{D_p}{D}, \tag{21}$$

where $D$ denotes the fractal dimension of a network of cities, and $D_p$ represents the average value of the fractal dimensions of the urban population distribution of different cities within the city network(Chen, 2014a). The reciprocal relation $q=1/p$ has been clarified above, the numerical relations between $p$, $q$ and $D$, $D_p$ should be explained in a few words. According to the principle of dimension consistency (Mandelbrot, 1982; Lee, 1989; Takayasu, 1990), we have a geometric measure relation as follows $N_m = \mu P_m^{(-D/D_p)}$, where $\mu$ is a proportionality coefficient. Comparing this equation with equation (10) yields $p = D/D_p$. Substituting equation (21) into equation (20) yields

$$b = D_p. \tag{22}$$

In fact, in terms of the general geometric measure relation (Chen, 2015; Feder, 1988; Lee, 1989; Mandelbrot, 1982; Takayasu, 1990), equation (14) can be rewritten as

$$P_m^{1/D_p} \propto L_m^{1/D_l}, \tag{23}$$

where $D_p$ refers to the dimension of $P_m$, and $D_l$ to the dimension of $L_m$. The symbol "$\propto$" denotes "be proportional to". As $L_m$ is a distance measurement, the dimension $D_l=1$. Thus equation (14) can be transformed into the following expression

$$P_m = \kappa L_m^{D_p/D_l} = \kappa L_m^{D_p}. \tag{24}$$

Comparing equation (24) with equation (14) shows $\sigma = D_p$. In light of equation (18), we have $b = D_p$, which is just equation (22). This suggests that the distance exponent of the gravity model is the average fractal dimension of urban populations of cities inside a geographical region. In empirical studies, it is hard to calculate $D_p$, but it is easy to evaluate $q$ and $D$, and thus the distance exponent can be estimated by equation (20).

In practice, the distance exponent can be estimated by the Minkowski–Bouligand dimension of a spatial distribution and the Zipf exponent of the corresponding rank-size distribution. The Minkowski–Bouligand dimension is often termed the box-counting dimension or box dimension by usage (Schroeder, 1991). The Zipf exponent can be given by the two-parameter Zipf model, equation (5). The box dimension is always defined as below

$$D_b = -\lim_{\varepsilon \to 0} \frac{\ln N(\varepsilon)}{\ln \varepsilon}, \tag{25}$$

where $\varepsilon$ refers to the linear scale of boxes, $N(\varepsilon)$ to the number of the nonempty boxes, and $D_b$ is the box dimension. Therefore, the box dimension can be evaluated by the following relations

$$N(\varepsilon) = N_1 \varepsilon^{-D_b}, \tag{26}$$

where $N_1$ is the proportionality coefficient. Theoretically, $N_1=1$, and empirically, $N_1$ is close to 1.

Thus the distance exponent is given by

$$b = qD_b, \qquad (27)$$

which is a substituted form of equation (20).

It is necessary to clarify the similarities and differences between these methods. Three scaling approaches for estimating the distance exponent can be illustrated as follows (Figure 1). The first approach is direct calculation using size-distance scaling, equation (24), and the fractal dimension of size measurement, $D_p$, is just the distance exponent ($b=D_p$). The second approach is indirect estimation using number-size scaling and number-distance scaling, that is, equations (10) and (12). These combine into equation (20), and the product of the fractal dimension of network, $D$, and the reciprocal of the scaling exponent, $p$, is theoretically equal to the distance exponent ($b=D/p$). The third approach is indirect estimation using Zipf's law and the box-counting method including equations (5) and (26), which combine into equation (24). The product of the Zipf exponent, $q$, and the box dimension, $D_b$, is approximately equal to the distance exponent ($b=qD_b$).

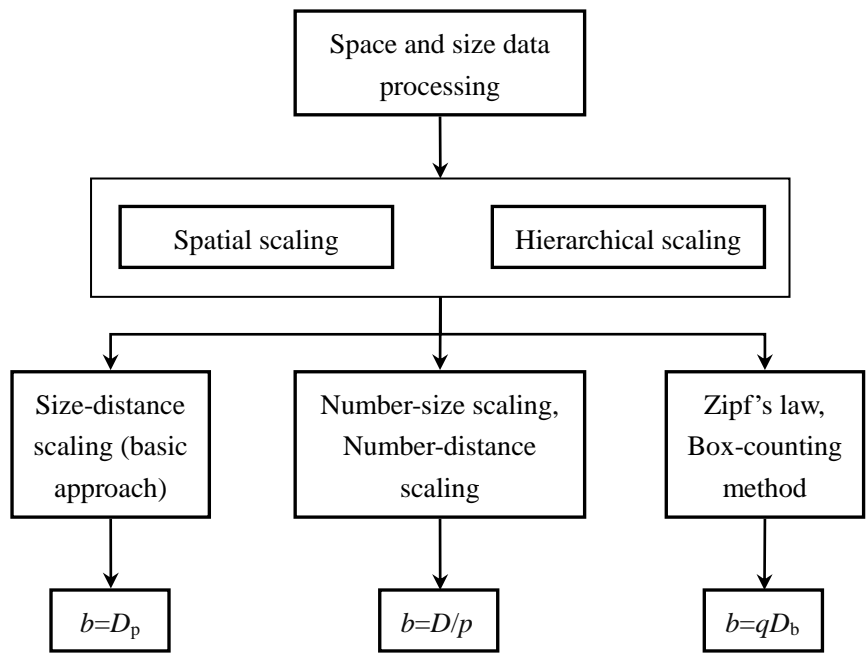

**Figure 1 Three approaches based on spatial and hierarchical scaling for estimating the distance exponent of the gravity model**

# 3 Materials and methods

## 3.1 Study area, materials and methods

Chinese cities can be employed to test the theoretical results and to show how to estimate the distance exponent of the gravity model. The study area is the whole mainland China. Two datasets of city sizes are available including the observations of the fifth census (2000) and the sixth census (2010). The numbers of cities are 666 in 2000 and 654 in 2010. The 2000-year dataset was processed and organized by Zhou and Yu (2004a; 2004b). The analytical procedure is as follows. **Step 1**: **examine city rank-size patterns**. The rank-size distribution of cities indicates an urban hierarchy with cascade structure (Chen, 2012a; Chen, 2012b). Generally speaking, only the cities with size larger than a threshold value comply with the rank-size rule. Smaller cities can be regarded as outliers beyond the scale-free range (Chen, 2015). By means of a double logarithmic plot and residual outlier analysis, we can determine a scaling range for the rank-size distribution. The scaling range appears as a straight line segment on the log-log plot. **Step 2**: **reconstruct order space**. Hierarchical structure suggests an order-space of urban systems, which corresponds to the real space of network structure (Chen, 2014a). If a system of cities follows Zipf's law, it can be converted into a self-similar hierarchy of cities with a number of levels. The rank will be replaced by city number, and the city size will be replaced by average size of each level. Thus the order-space based on hierarchical structure will substitute for the rank-size order space based on the rank-size distribution. **Step 3**: **measure the nearest neighboring distance (NND).** In each level of an urban hierarchy, each city possesses coordinate cities (Ye *et al*, 2000). The distance between a city and its nearest coordinate city is termed the *nearest neighboring distance* (Clark and Evans, 1954; Rayner *et al*, 1971), which is a basic measurement for studies on central places and self-organized networks of cities (Chen, 2011). **Step 4**: **build models and estimate model parameters.** Using the datasets of hierarchies of cities, we can make fractal models, allometric scaling models, and hierarchical scaling models based on the hierarchical order-space. The ordinary least-squares (OLS) method can be adopted to evaluate the fractal dimension and the related scaling exponents because this algorithm has its significant merits (Chen, 2015). Sometimes however, the OLS algorithm is not the best approach for evaluating a power exponent. A new method based on the maximum likelihood estimation (MLE) is proposed by Clauset *et al* (2009) to identify various power-law distributions and to estimate scaling exponents. This algorithm combines the methods of maximum-likelihood fitting with goodness-of-fit tests based on the Kolmogorov-Smirnov statistic and likelihood ratios. Unfortunately, the MLE-based approach cannot be applied to our datasets in this study for the following reasons. First, the MLE-based method is suitable for binned data, while this study is on pairs of cascade sequences. Second, the MLE-based method is developed for power-law frequency distributions, while this study is on fractal measure relations.

### 3.2 Calculations

The calculation procedure comprises four steps (Two Supplementary Materials are provided to show how to deal with the data and estimate the parameters, see Files S1 and S2). **The first step is to investigate rank-size patterns.** The common two-parameter Zipf model cannot be well fitted to the datasets of China's city sizes. In fact, the rank-size distribution of Chinese cities should be modeled with the three-parameter Zipf's model (Chen, 2016; Gabaix, 1999; Gell-Mann, 1994; Mandelbrot, 1982; Winiwarter, 1983). The model can be expressed as

$$P(k) = C(k+\varsigma)^{-q}, \qquad (28)$$

where $k$ refers to city rank ($k$=1, 2, 3,…), and $P(k)$ to the size of the $k$th city. As for the parameters, $C$ is a proportionality coefficient, $q$ denotes the Zipf scaling exponent, and $\varsigma$ represents a scale-translational factor of city rank. The largest 550 cities are inside the scaling ranges, and the three-parameter rank-size patterns are displayed in Figure 2. The first four or five ranks are removed due to the incomplete structure of Chinese hierarchies of cities (Chen, 2016). What is more, the cut-off point of scaling range is identified by combining eye observation, goodness of fit with residual outlier analysis (Chen, 2015).

**The second step is to reconstruct the order-space of the urban system.** The three-parameter Zipf model suggests an absence of the leading cities in one or more top levels of a hierarchy of cities (Chen, 2016). For 2000-year cities, the scale-translational parameter $\varsigma$=5, which suggests the first and second levels are absent. For the 2010-year cities, the adjusting parameter $\varsigma$=4; this also suggests the absence of the first and second levels in the hierarchy. If the number ratio of different levels of cities is $r_n$=2, cities can be organized into urban hierarchies by year (Table 1). Each hierarchy of cities comprises 8 levels, which reflects the order-space of the corresponding network of cities ($m$=3, 4, …, 10). The first and second levels are vacant, and the last level is a lame-duck class, in which the real number of cities is smaller than the expected number (Davis, 1978).

**Table 1 Numbers, average sizes, and average least-distances of China's cities**

| Order | 2000 | | | 2010 | | |
|---|---|---|---|---|---|---|
| $m$ | Number $N_m$ | Size $P_m$ | Distance $L_m$ | Number $N_m$ | Size $P_m$ | Distance $L_m$ |
| 1 | 1 | -- | -- | 1 | -- | -- |
| 2 | 2 | -- | -- | 2 | -- | -- |
| 3 | 4 | 890.8895 | 931.7267 | 4 | 1319.619 | 981.3669 |
| 4 | 8 | 446.1617 | 563.9267 | 8 | 697.352 | 606.1919 |
| 5 | 16 | 233.6337 | 534.0708 | 16 | 340.611 | 494.4720 |
| 6 | 32 | 123.1711 | 212.2135 | 32 | 170.434 | 254.0088 |
| 7 | 64 | 68.3960 | 153.5515 | 64 | 81.168 | 167.1989 |
| 8 | 128 | 35.1952 | 132.9058 | 128 | 44.626 | 129.5111 |
| 9 | 256 | 17.8514 | 98.7091 | 256 | 22.160 | 99.6486 |
| 10 | 158 | 8.6543 | 133.8155 | 146 | 9.900 | 135.3284 |

**Note**: The distance unit is kilometer (km). The two top levels ($m$=1, 2) are absent according to the three-parameter Zipf model. The last level ($m$=10) represents a lame-duck class because of undergrowth of cities. The principle and procedure of creating the hierarchy of cities in this table are illuminated by Chen (2016).

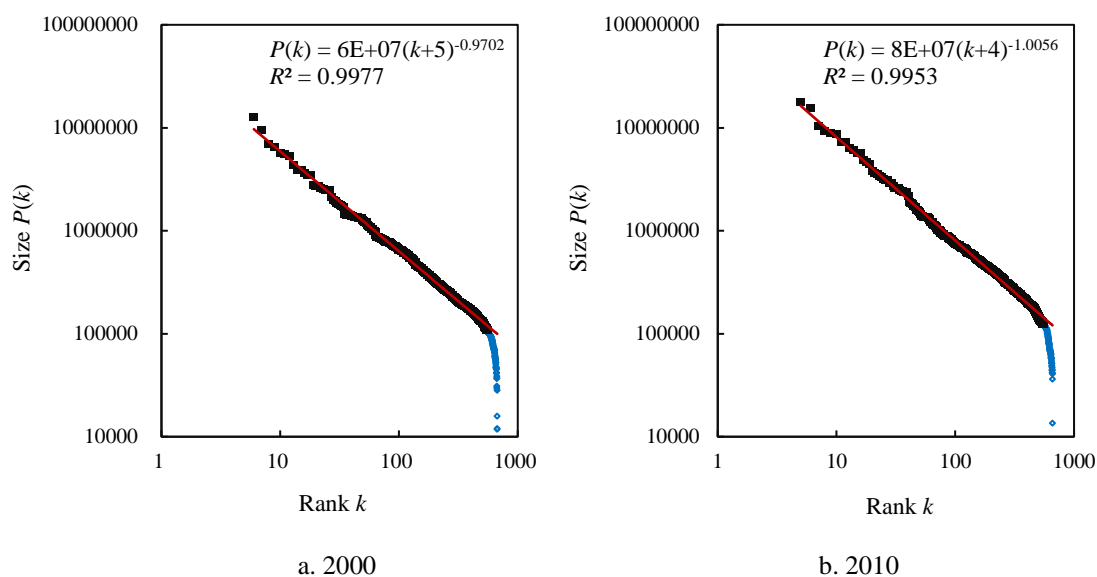

a. 2000　　　　　　　　　　　　　　b. 2010

**Figure 2 The rank-size distributions of China's cities based on the three-parameter Zipf's model**

[**Note:** A three-parameter Zipf distribution corresponds to an incomplete hierarchy of cities, on the top of which one or more levels have not developed.]

**The third step is to measure the nearest neighboring distance.** For a given level in an urban hierarchy, each city has a nearest neighbor. Based on the hierarchy with 8 levels, the distances between different cities can be measured by Arc GIS. For each level, the distances can be arranged as a distance matrix, from which we can extract vectors of the nearest neighboring distances. The average values of the elements in the $m$th vector represent the nearest neighboring distance of the $m$th level ($m$=3, 4, …, 10). In fact, for medium-sized and small cities, the average value of coordination city numbers is close to 6 (Haggett, 1969; Niu, 1992; Ye *et al*, 2000). In order to lessen the negative effect of random fluctuation of city distributions, we can take the next nearest neighbor of a city into account and measure the *second nearest neighboring distance* (SNND). After extract the NND data, we can further extract the SNND from each distance matrix. The mean of NND and SNND of each level can be termed *dual average neighboring distance* (DAND) (Table 2). In this study, the DAND values are employed to carry out the spatial modeling and analysis.

**Table 2 Average distances of China's cities to nearest neighbors and second nearest neighbors**

| $m$ | 2000 | | | 2010 | | |
|---|---|---|---|---|---|---|
| | NND | SNND | DAND | NNDs | SNND | DAND |
| 3 | 807.4446 | 1056.0087 | 931.7267 | 591.5387 | 1371.1950 | 981.3669 |
| 4 | 336.6311 | 791.2222 | 563.9267 | 347.4975 | 864.8864 | 606.1919 |
| 5 | 444.4505 | 623.6911 | 534.0708 | 412.1159 | 576.8281 | 494.4720 |
| 6 | 154.5868 | 269.8403 | 212.2135 | 190.2816 | 317.7361 | 254.0088 |
| 7 | 129.6058 | 177.4973 | 153.5515 | 131.7646 | 202.6332 | 167.1989 |
| 8 | 95.9974 | 169.8142 | 132.9058 | 99.5365 | 159.4856 | 129.5111 |
| 9 | 81.4986 | 115.9197 | 98.7091 | 82.0976 | 117.1997 | 99.6486 |
| 10 | 103.4494 | 164.1817 | 133.8155 | 110.0906 | 160.5662 | 135.3284 |

**Note**: The distance unit is kilometer (km). NND refers to the distance of a city to its nearest neighbor, SNND to the distance of a city to its second nearest neighbor. In this table, both NND and SNND represent the mean of a level of cities, and DAND denotes the dual average value of NND and SNND.

**The fourth step is to build models using the hierarchical datasets.** Except for the lame-duck class, the city numbers, average city sizes, and the average distances follow exponential laws. City numbers increase exponentially, and average city sizes and average distances decrease exponentially over order $m$. From the three exponential laws, a set of power laws follows, including the distance-number scaling relation, the distance-size scaling relation, and the number-size scaling relation. The hierarchical scaling relationship between city numbers and average city sizes of different levels is equivalent to the three-parameter Zipf model, and the scaling exponent $q$ is theoretically equal to the Zipf exponent (Figure 3). The hierarchical scaling relationship between average distances and city numbers is a fractal model of an urban network, and the fractal dimension $D$ is just the fractional dimension of a central-place system (Figure 4). The hierarchical scaling relationship between average distances and average city sizes is equivalent to the allometric scaling relationship between city sizes and the area of service regions of cities, and the scaling exponent $\sigma$ is just the average fractional dimension of city population distributions (Figure 5). As indicated above, in theory, the parameter $\sigma$ can be estimated by the product of $q$ and $D$ (Table 3).

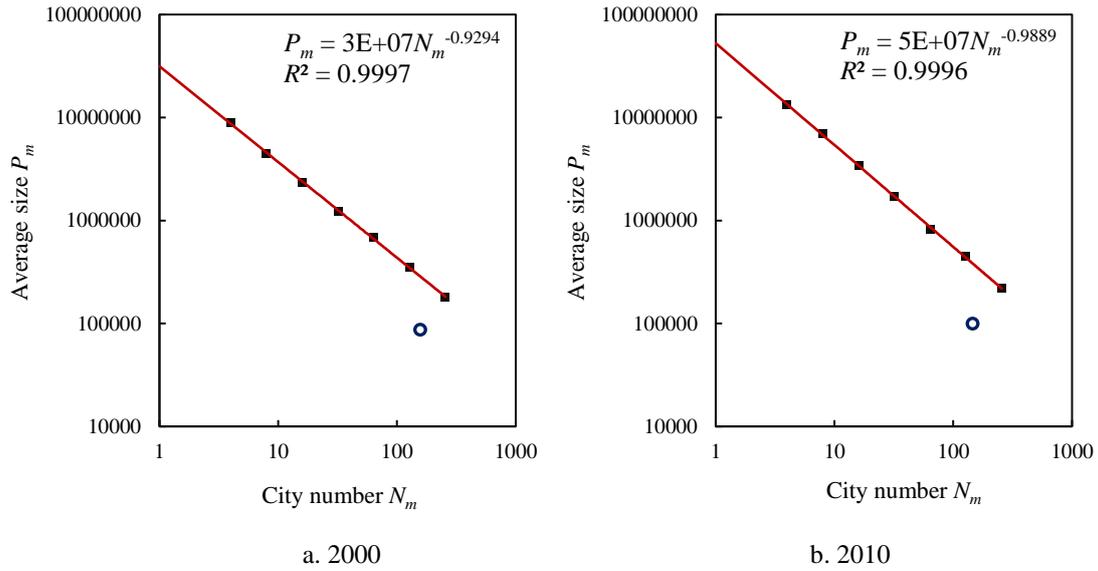

a. 2000   b. 2010

**Figure 3 The hierarchical scaling relationships between the numbers of China's cities and the corresponding average city sizes [Note:** Figure 3 corresponds to Figure 2. The small circles represent outliers indicative of the lame-duck classes.**]**

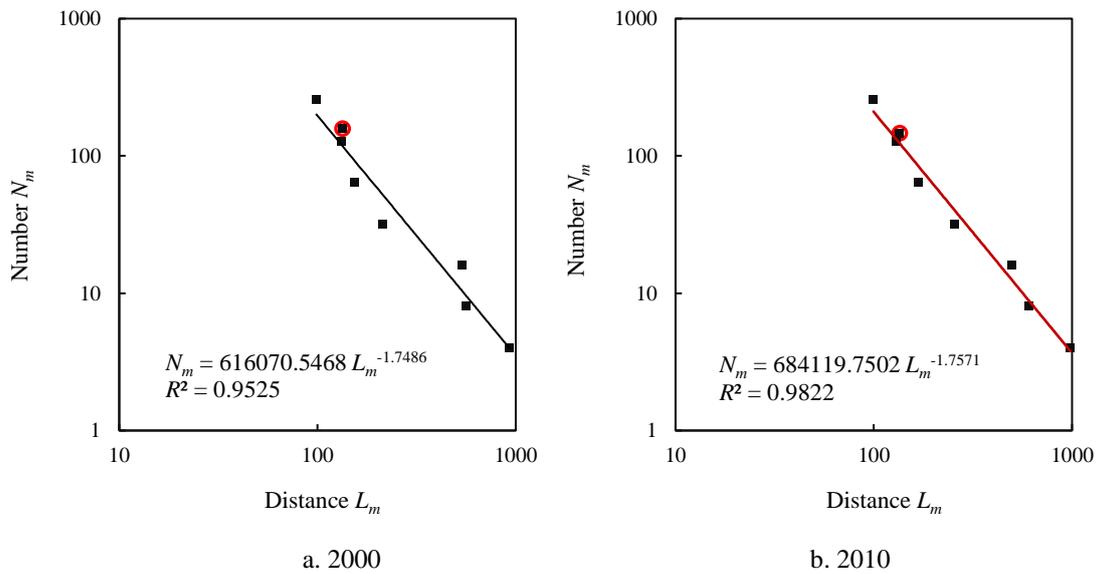

a. 2000   b. 2010

**Figure 4 The hierarchical scaling relationships between numbers of China's cities and the**

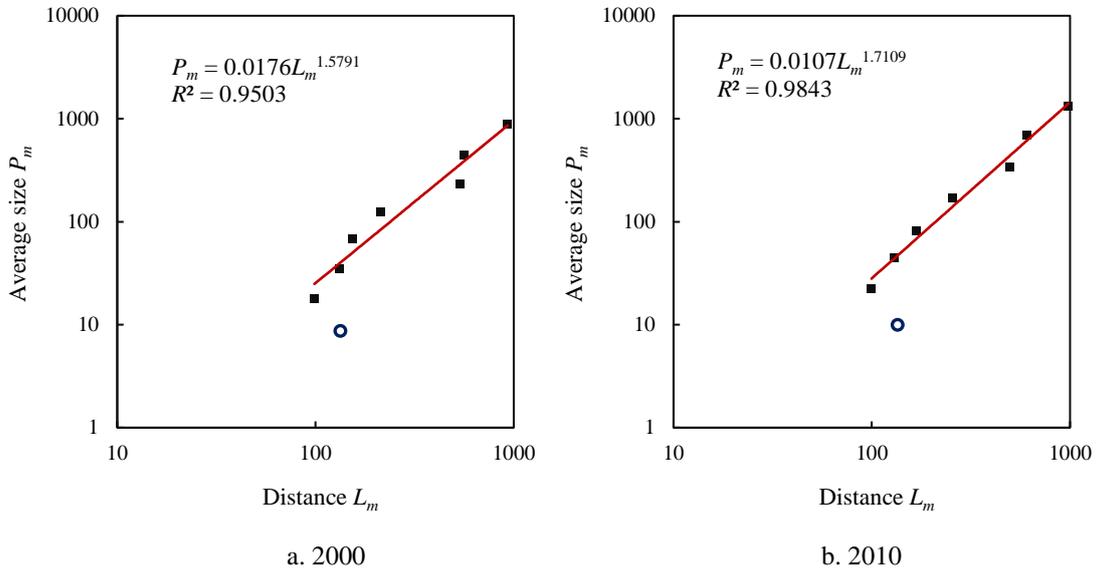

a. 2000    b. 2010

**Figure 5 The hierarchical scaling relationships between the minimum distances of China's cities and the corresponding average city sizes [Note:** The small circles represent outliers indicative of the lame-duck classes.]

### 3.3 Analysis

Two approaches have been used to estimate the distance exponent of the gravity model of China's cities. One is to use the distance-size hierarchical scaling, and the other is to utilize both the distance-number hierarchical scaling and the number-size hierarchical scaling. If we don't remove the bottom level (lame-duck class), the error between the direct result ($b=\sigma$ value) and the indirect result ($b \approx qD$ value) will be large. If we treat the bottom level as an outlier and eliminate it, the directly estimated value will be close to the indirectly estimated value, that is, $qD \approx \sigma$ (Table 3). This suggests that the outlier of a dataset can cause significant deviation in parameter estimation. Intuitively, the distance exponent value should have decreased because of development of traffic technology from 2000 to 2010 year. However, the distance exponent actually increased during this decade due to intervening opportunities resulting from city development (as for intervening opportunity, see Stouffer, 1940).

**Table 3 The estimated values of model parameters of the hierarchies of China's cities based on all**

data points and scaling ranges

| Scale | Parameter | 2000 | | 2010 | |
|---|---|---|---|---|---|
| | | Estimated value | $R^2$ | Estimated value | $R^2$ |
| All data points (Including bottom level) | $q$ | 1.0355 | 0.9403 | 1.1047 | 0.9314 |
| | $D$ | 1.7486 | 0.9525 | 1.7571 | 0.9822 |
| | $\sigma$ | 1.7774 | 0.8631 | 1.9199 | 0.8950 |
| | $qD$ | 1.8106 | -- | 1.9411 | -- |
| Scaling range (Eliminating bottom level) | $q$ | 0.9294 | 0.9997 | 0.9889 | 0.9996 |
| | $D$ | 1.7008 | 0.9527 | 1.7280 | 0.9823 |
| | $\sigma$ | 1.5791 | 0.9503 | 1.7109 | 0.9843 |
| | $qD$ | 1.5808 | -- | 1.7088 | -- |

Another factor that impacts the estimated value of the scaling exponent is the spatial pattern of cities. The spatial distribution of urban populations is often a self-affine pattern rather than a self-similar pattern. In other words, the urban population distribution is based on anisotropic growth instead of isotropic growth. As a result, the points on a double logarithmic scatter plot form two trend lines rather than one. The scaling break used to be treated as bi-fractals (White and Engelen, 1993; White and Engelen, 1994). In many cases, bi-fractals proceed from self-affine distributions. The spatial distributions of Chinese cities bear significant self-affinity (Figure 6). However, they evolve from self-affine patterns into self-similar patterns because the difference between the fractal dimension values of the two scaling ranges became smaller from 2000 to 2010 (Table 4). In order to avoid the influence of spatial self-affinity, the fractal dimension can be estimated by the box-counting method, which yields what is called box dimension mentioned above. The box-counting method has been employed to estimate the fractal dimension of urban forms (Benguigui *et al*, 2000; Feng and Chen, 2010; Shen, 2002). For simplicity, the classical box-counting method can be replaced by the functional box-counting method (Chen, 1995; Lovejoy *et al*, 1987). This method is simple and effective. For example, for the cities of Henan province of China in 2000, the Zipf exponent of the city-size distribution was about $q$=0.968 (Jiang and Yao, 2010), and the box dimension of the central-place network was about $D_b$=1.8585(Chen, 2014b). Thus the distance exponent of the gravity model of Henan's urban system was about $b$=0.968*1.8585=1.799 in 2000. As far as mainland China is concerned, in 2000 the box dimension of the urban network is about $D_b$=1.8068, the Zipf exponent is $q$=0.9702, thus $b$=0.9702*1.8068=1.7530; in 2010 $D_b$=1.8193, $q$=1.0056, and we have $b$=1.0056*1.8193=1.8295 [see Chen (2014b) and Feng and Chen (2000) for

the box counting method].

**Table 4 Fractal dimension values of population distributions based on bi-scaling ranges**

| Item | 2000 | | 2010 | |
|---|---|---|---|---|
| | Parameter | $R^2$ | Parameter | $R^2$ |
| Scaling range 1 ($L_m \leq L_6$) | 2.5915 | 0.9762 | 2.1669 | 0.9898 |
| Scaling range 2 ($L_m \geq L_6$) | 1.2944 | 0.8814 | 1.5474 | 0.9597 |
| Intersection ($L_6$) | 254.0088 | | 212.2135 | |
| Average of two parameters | 1.9430 | | 1.8572 | |
| Difference of two parameters | 1.2971 | | 0.6195 | |

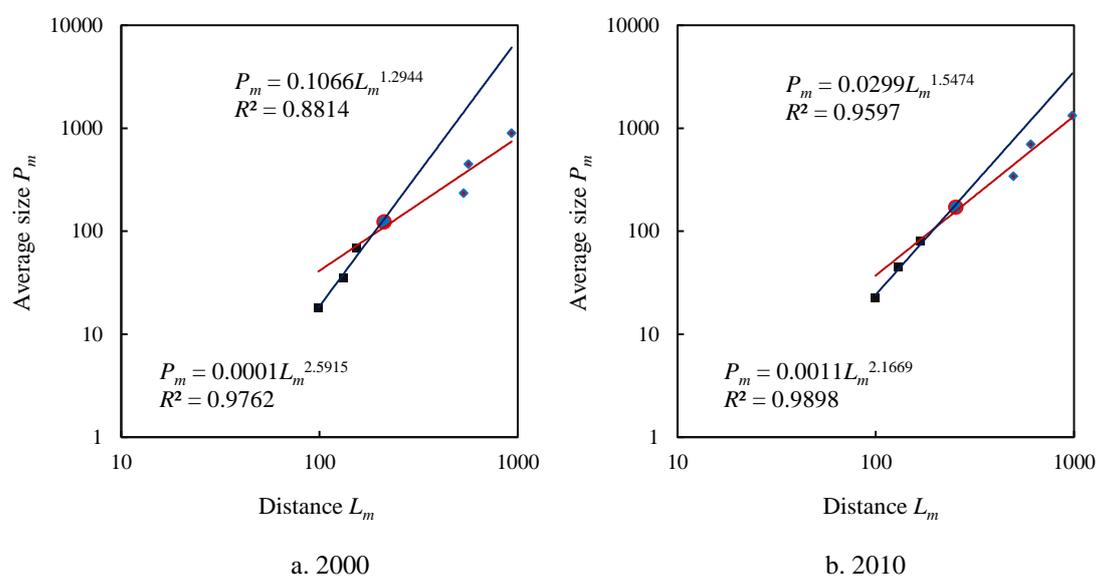

a. 2000      b. 2010

**Figure 6 Self-affine patterns and the bi-scaling relationships between the minimum distances of China's cities and the corresponding average city sizes [Note:** The small circles represent the intersections of the two scaling ranges**]**

## 4 Discussion

### 4.1 Evaluation of different methods

The research results support the theoretical inference that the rank-size scaling and fractal modeling can be employed to evaluate the distance exponent of the gravity model. On the basis of the empirical analyses, a preliminary evaluation of the three approaches to estimating the distance exponent can be made. The first approach based on size-distance scaling is simple and direct, but it is sometimes hindered by self-affine spatial distributions. The second approach based on the combination of number-distance scaling and number-size scaling is effective, but it is indirect and sometimes influenced by spatial self-affinity. The third approach, the combination of Zipf's law and the box-counting method, can avoid the negative influence of self-affine patterns, but it is difficult to match the sample of a size distribution with that of the corresponding spatial distribution. Nothing is perfect. Each method has its merits and defects. A comparison of advantages and disadvantages between these approaches is displayed in Table 5.

Table 5 A comparison of three approaches for estimating the distance exponent

| Method | Model | Formula | Merit | Defect |
|---|---|---|---|---|
| The first approach | $P_m = \kappa L_m^{D_P}$ | $b = D_P$ | Direct and simple | Sensitive to self-affine patterns |
| The second approach | $N_m = \pi L_m^{-D}$, $N_m = \mu P_m^{-p}$ | $q = R^2 / p$, $b = qD$ | Easy to understand | Indirect and influenced by self-affinity |
| The third approach | $P(k) = C(k+\varsigma)^{-q}$, $N(\varepsilon) = N_1 \varepsilon^{-D_b}$ | $b = qD_b$ | Easy to calculate | Indirect and influenced by sample difference |

The gravity model has been employed to research spatial interactions in both natural and human geographical phenomena. In empirical studies, there are two basic distance-decay functions for modeling gravity (Haggett *et al*, 1977): one is the inverse power law (Kang *et al*, 2012; Jung *et al*, 2008), and the other, the negative exponential law (Balcan *et al*, 2009; Wilson, 2000). If we examine the local or partial spatial interactions between different places, the gravity models take on generalized production functions such as (Lee *et al*, 2014; Machay, 1958)

$$I_{ij} = K \frac{P_i^\alpha P_j^\beta}{L_{ij}^\gamma}, \tag{29}$$

which is based on the power-law decay, and (Balcan *et al*, 2009)

$$I_{ij} = K \frac{P_i^\alpha P_j^\beta}{e^{\gamma L_{ij}}}, \tag{30}$$

which is based on exponential decay. In these formulae, $K$, $α$, $β$, and $γ$ are parameters. However, if all the spatial relationships in a network are taken into consideration, equations (29) and (30) can be transformed into the standard forms as follows (Chen, 2015)

$$I_{ij} = G\frac{P_i P_j}{L_{ij}^b}, \tag{31}$$

$$I_{ij} = G\frac{P_i P_j}{e^{bL_{ij}}}, \tag{32}$$

where the notation is the same as equations (1) and (2). The parameter *b* in equation (32) is a distance-decay coefficient. Different types of gravity models have different spheres of application. Generally speaking, the exponential-based model is more applicable to smaller regions or simpler systems while the power-based model is more suitable for larger regions or more complex systems (Chen, 2015). If the distance-decay functions are appropriately selected, the gravity models can be well fitted to the observational data of spatial flows (e.g., Balcan *et al*, 2009; Batty and Karmeshu, 1983; Kang *et al*, 2012; Goh *et al*, 2012; Jung *et al*, 2008; Lee *et al*, 2014). From gravity modeling, we can gain new insights into the dynamic process of spatial interactions.

**4.2 Further thinking**

Using the mathematical relations derived above, we can develop the gravity theory of human geography and solve the difficult problems which puzzle geographers for a long time. One problem is the question of the gravity constant. During the quantitative revolution of geography (1953-1976), geographers tried in vain to find constant values for the parameters of the gravity model, *G* and *b*, by means of mathematical derivation and statistical analysis (Harvey, 1969). Since then, many geographers cast doubt on the theoretical basis of the gravity model. Today, we know that human geographical systems differ from classical physical systems. The natural laws of human geographical systems don't comply with the rule of spatio-temporal translational symmetry (Chen, 2015). Thus we cannot find constant values for the model parameters of geographical gravity. As indicated above, the distance exponent of the urban gravity model is the average fractal dimension of urban population of the cities within a geographical region. The rest may be deduced by analogy. However, one basic property of geographical systems is regional differentiation. The condition of one geographical region differs from those of other geographical regions. Thus the average fractal dimension of urban population of one region is different from that of another region.

Another problem is the dimension dilemma. According to dimensional analysis, the theoretical value of the distance exponent of the gravity model should equal 1, 2, or 3, which are dimension values of Euclidean geometry. However, the observed values from empirical studies often deviated from theoretical values significantly and took on fractional numbers (Haynes, 1975; Haggett *et al*, 1977; Rybski *et al*, 2013). Today, it is easy to solve the difficult problem of dimensional analysis. The distance exponent is, in essence, a fractal dimension of urban population (Chen, 2015). A fractal dimension always exhibits a fractional number. So it is not surprising that empirical results of distance exponent estimation fail to equal 1, 2, or 3. If we define our study area within a 2-dimensional space, the distance exponent will range from 1 to 2; if we definite our study area inside a 3-dimensional space, the distance exponent will vary from 1 to 3. Generally speaking, the distance exponent falls between 1 and 3 in empirical studies. This lends further support to the suggestion that fractal geometry, allometric analysis, and network theory can be integrated into a new theoretical framework to reinterpret many of our theories in physical and human geography (Batty, 1992; Batty, 2008; Chen, 2015).

It has not been reported in the previous literature that Zipf's law and the fractal dimension of urban network can be combined to estimate the distance exponent of the gravity model. This is just the novelty of this work. However, this research has two main shortcomings. First, the case study is only based on the census data of Chinese cities. We have no data for cities in other countries for the time being. In this case, we cannot carry out a cross-sectional comparative analysis. Second, the methods are developed for self-similar fractal structures. However, many urban systems are random self-affine fractals. We have no effective approaches for dealing with the self-affine spatial distribution of cities. Anyway, the value of a paper rests with its inspiration rather than its perfection.

## 5 Conclusions

By the empirical study based on Chinese cities, we can reconsider the geographical gravity model and its parameters. New findings lie in three aspects, that is, the fractality and scaling behind the gravity model, the fractal dimension property of the distance exponent, and the effective methods of parameter estimation. All these are based on the hierarchy of cities with cascade structure. The main conclusions can be reached as follows. **First, the geographical gravity model is a fractal model of spatial interaction associated with a self-similar hierarchy.** Before the advent of fractal geometry, the gravity model was explained by ideas from Euclidean geometry. However, geographical systems are fractal systems with fractional dimensions. Many difficult problems such as the dimensional dilemma and gravity constants resulted from traditional explanations based on Euclidean geometry. Today, many theoretical problems relating to the gravity model and its parameters can be easily solved using ideas from fractals, scaling, and allometry based on a hierarchical structure. **Second, the distance exponent of the gravity model is a kind of fractal dimension indicating a given geographical quantity of matter.** The property of fractal dimension depends on size measurements such as population size, urban area, and economic product. If we use population sizes of cities to determine the attraction power, the distance exponent will represent the average fractal dimension of the urban populations of cities within a geographical region. Similarly, if we employ urban areas to define the attraction power, the distance exponent will reflect the average fractal dimension of urbanized areas of cities in a geographical region. The rest may be deduced by analogy. **Third, the distance exponent of urban gravity can be estimated by Zipf's law and the box-counting method.** The distance exponent equals the product of the hierarchical scaling exponent of size distributions of cities and the fractal dimension of a network of cities in a study area. It is easy to evaluate the scaling exponent of a rank-size distribution (indicative of a hierarchy) and corresponding fractal dimension of an urban system (indicative of a network). The rank-size exponents can be computed by Zipf's law, and the fractal dimensions of urban networks can be calculated using the box-counting method. Thus the distance exponent can be estimated by using fractal modeling of size distributions and spatial distributions of cities.

## Acknowledgements

This research was sponsored by the National Natural Science Foundation of China (Grant No. 41671167). The supports are gratefully acknowledged.

# Appendixes

## Appendix 1 The relationships between Zipf's law, Pareto distribution, and hierarchical scaling law

From Zipf's law, we can derive Pareto distribution and hierarchical scaling law. For simplicity, let's see the rank-size distribution based on pure Zipf's law, $P(k)=1/k$. Using this equation, we can generate a harmonic sequence such as 1, 1/2, 1/3, …, 1/k, …. According to the generalized Davis's $2^n$ rule, the harmonic sequence can be organized into a hierarchy with cascade structure (Table I). From the hierarchy, we can derive equation (6) and equation (10). The city number in each class is $N_m$=1, 2, 4, …, $2^{m-1}$, where $m$=1,2,3,…. This sequence can be formulated as

$$N_m = 2^{m-1} = N_1 r_n^{m-1},\qquad(A1)$$

in which $N_1$=1 and $r_n$=2. Let the size threshold be $P$=1, 1/2, 1/4,…, $1/2^{m-1}$. Clearly, the number of cities with population size greater than or equal to $P$ is $N(P)$ =1,2,4,…, $2^{m-1}$. Thus we have

$$N(P) = 2^{m-1} = P^{-1} = \eta P^{-p},\qquad(A2)$$

where $\eta$=1 and $p$=1. Equation (A2) is equivalent to the standard Pareto distribution. Comparing equation (A1) with equation (A2) shows $N_m$=$N(p)$. On the other hand, it can be proved that the total size of cities at each level is $S_m$→ln(2)=0.6931 (Chen, 2012a). The average city size of each level is

$$P_m = \frac{S_m}{N_m} = \frac{\ln(2)}{2^{m-1}} = P_1 r_p^{1-m},\qquad(A3)$$

in which $P_1$=ln(2) and $r_p$=2. From equation (A3) it follows

$$N_m = \frac{S_m}{P_m} = \ln(2) P_m^{-1} = \mu P_m^{-p},\qquad(A4)$$

where $\mu$→ln(2) and $p$→1. Equation (A4) is equivalent to the pure Zipf's law. The scaling exponent $p$ is asymptotically close to 1. This mathematical process can be generalized to common Zipf's law and Pareto distribution (A Supplementary Material is provided to show how to transform the harmonic sequence into a hierarchy, see File S3).

Table I A hierarchy of 2047 cities with cascade structure coming from pure Zipf's distribution

| Level $m$ | City number $N_m$, $N(P)$ | Hierarchy (The first number at each level acts as the size threshold, $P$) | | | | | Sum $S_m$ | Average size $P_m$ |
|---|---|---|---|---|---|---|---|---|
| 1 | 1 | **1** | | | | | 1 | 1 |
| 2 | 2 | **1/2** | 1/3 | | | | 0.8333 | 0.4167 |
| 3 | 4 | **1/4** | 1/5 | 1/6 | 1/7 | | 0.7595 | 0.1899 |
| 4 | 8 | **1/8** | 1/9 | 1/10 | 1/11 | …… | 0.7254 | 0.0907 |
| 5 | 16 | **1/16** | 1/17 | 1/18 | 1/19 | …… | 0.7090 | 0.0443 |
| 6 | 32 | **1/32** | 1/33 | 1/34 | 1/35 | …… | 0.7010 | 0.0219 |
| 7 | 64 | **1/64** | 1/65 | 1/66 | 1/67 | …… | 0.6971 | 0.0109 |
| 8 | 128 | **1/128** | 1/129 | 1/130 | 1/131 | …… | 0.6951 | 0.0054 |

| 9 | 256 | **1/256** | 1/257 | 1/258 | 1/259 | …… | 0.6941 | 0.0027 |
| 10 | 512 | **1/512** | 1/513 | 1/514 | 1/515 | …… | 0.6936 | 0.0014 |
| 11 | 1024 | **1/1024** | 1/1025 | 1/1026 | 1/1027 | …… | 0.6934 | 0.0007 |

**Note**: Starting from the fourth level, only the first four numbers are displayed due to the limited space. If the city number is not equal to $2^{m-1}$, the final level will be a lame-duck class.

## Appendix 2 The principle of dimension consistency and fractal measure relation

As early as the ancient Greek times, mathematicians found the principle of dimension consistency. That is, a measure is in proportion to another measure if and only if the two measure have the same dimension. For length $L$, area $A$, volume $V$, and arbitrary measure $M$, we have a proportion relation $L^{1/1}=k_a A^{1/2}=k_v V^{1/3}=k_m M^{1/d}$, where $k_a$, $k_v$, $k_m$ are proportionality coefficients, and $d$ is the dimension of $M$ (Lee, 1989; Feder, 1988). This is well-known geometric measure relation that can generalized to fractal measure relation (Feder, 1989; Mandelbrot, 1982; Takayasu, 1990). Equation (16) can be obtained by the dimensional analysis based on the geometric measure relation. It is easy to derive equation (14) from equations (8) and (9). Replacing $m$ by $m+1$ in equation (14) yields

$$P_{m+1} = \kappa L_{m+1}^\sigma, \tag{B1}$$

Combining equation (B1) with equation (14) yields

$$\frac{P_m}{P_{m+1}} = \left(\frac{L_m}{L_{m+1}}\right)^\sigma, \tag{B2}$$

which shares the same mathematical structure with equation (4). According to the principle of dimension consistency, $P_m$, $P_{m+1}$, $P_i$, and $P_j$ have the same space dimension $D_p$, and $L_m$, $L_{m+1}$, $L_{ix}$, and $L_{jx}$ bear the same space dimension $D_l$. Thus we have two geometric measure relations as follows

$$\frac{P_m}{P_{m+1}} = \left(\frac{L_m}{L_{m+1}}\right)^{D_p/D_l}, \tag{B3}$$

$$\frac{P_i}{P_j} = \left(\frac{L_{ix}}{L_{jx}}\right)^{D_p/D_l}. \tag{B4}$$

Comparing equations (B3) and (B4) with equations (B2) and (4) yields

$$\sigma = \frac{D_p}{D_l} = b. \tag{B5}$$

Substituting equation (B5) into equation (B2) yields equation (16), immediately. This suggests that, although $P_m$ and $P_{m+1}$ differ from $P_i$ and $P_j$ and $L_m$ and $L_{m+1}$ differ from $L_{ix}$ and $L_{jx}$, the scaling exponent based on $P_m/P_{m+1}$ and $L_m/L_{m+1}$ is equivalent to that based on $P_i/P_j$ and $L_{ix}/L_{jx}$.

# Legends of Supplementary Materials for on-line publication only

### S1 Data processing for Chinese hierarchy of cities in 2000 (XLSX)

Based on the 2000 year urban population census data of China, the nearest neighboring distance (NND) and second nearest neighbor distance (SNND) were calculated, and the rank-size series of urban population was transformed into a hierarchy with cascade structure according to the generalized $2^n$ rule. The distance matrixes were extracted by ArcGIS, and the other processes were fulfilled by MS Excel.

### S2 Data processing for Chinese hierarchy of cities in 2000 (XLSX)

Based on the 2010 year urban population census data, the NND and SNND were calculated, and the rank-size series of urban population was transformed into a self-similar hierarchy. Comparing the data processing for 2000 with that for 2010, the reader will understand the whole process of data processing in this paper. Then, readers can make computation by writing computer programs.

### S3 The process of transforming Zipf's distribution into self-similar hierarchy (XLSX)

A harmonic sequence consisting of 2047 data points was generated using the pure Zipf model. The sequence representing standard Zipf distribution was converted into a hierarchy with cascade structure.